\documentclass[prd,aps,showpacs,preprintnumbers,amssymb]{revtex4}
\usepackage{graphicx}
\usepackage{epsf}
\usepackage{amsmath}
\usepackage{epstopdf}
\usepackage{multirow}
\usepackage{bm}

\begin{document}
\title{%
\hfill{\normalsize\vbox{%
\hbox{}
 }}\\
{Electromagnetic trace anomaly in a generalized linear sigma model}}

\author{Amir H. Fariborz
$^{\it \bf a}$~\footnote[1]{Email:
 fariboa@sunyit.edu}}

\author{Renata Jora
$^{\it \bf b}$~\footnote[2]{Email:
 rjora@theory.nipne.ro}}

\affiliation{$^{\bf \it a}$ Department of Matemathics/Physics, SUNY Polytechnic Institute, Utica, NY 13502, USA}
\affiliation{$^{\bf \it b}$ National Institute of Physics and Nuclear Engineering PO Box MG-6, Bucharest-Magurele, Romania}

\date{\today}

\begin{abstract}
We build  the electromagnetic trace anomaly effective term for a generalized linear sigma model with two chiral nonets, one with a quark-antiquark structure, the other one with a four quark content. In the leading order of this framework, we study the decays into two photons of the lowest isosinglet scalar mesons.     We find that the direct inclusion  of underlying mixing among two- and four-quark components in the trace anomaly term is essential in order for the model prediction to agree with the available experimental data on decay width of $f_0(980)$ to two photons.     Consequently,  this sets a lower bound of 0.5  KeV on the decay with of $f_0(500)$ to two photons.

\end{abstract}
\maketitle

\section{Introduction}

The inverted mass spectrum of the low lying scalar mesons with respect to the pseudoscalar and vector ones is a long standing low energy QCD puzzle \cite{PDG} for which various
solutions have been proposed \cite{vanBev}-\cite{tsm} almost all of them dealing with the particular quark substructure of the scalar mesons. In a series of papers  \cite{Jora1}-\cite{Jora5} we proposed and studied in detail a generalized linear sigma model with two chiral nonets, one with a two quark substructure the other one with a four quark content. In this framework the physical scalar states were found to have a significant  admixture of two- and four-quark components, with those below 1 GeV  generally containing a larger  four-quark component compared to those above 1 GeV.

The generalized linear sigma model described in \cite{Jora1}-\cite{Jora5} contained,  besides the relevant terms pertaining to mass and interactions,  also a term that mocks up the gluon axial anomaly. An extra  term corresponding to the electromagnetic  axial anomaly was further introduced in \cite{Jora6} where the decays of the pseudoscalar mesons to two photons were computed and studied with a good agreement with the experimental data.
It seems then natural to extend this picture to include also the trace anomaly and analyze the decays of scalar mesons to two photons in the same context.

In section II we  briefly present our generalized linear sigma model followed by a derivation of the relevant term in the Lagrangian that leads to the correct electromagnetic
trace anomaly in section III. In section IV we give our numerical computation for the decay of $f_0(500)$ and $f_0(980)$ to two photons  and discuss  the results.

\section{Generalized linear sigma model}

  The model of interest is a generalized linear sigma model with two chiral nonets, one with a quark-antiquark structure $M$, the other one with a four quark structure $M'$:
\begin{eqnarray}
&&M=S+i\Phi
\nonumber\\
&&M'=S'+i\Phi',
\label{mode638456}
\end{eqnarray}
where $S$  and $S'$ represent the scalar nonets and $\Phi$ and $\Phi'$ the pseudoscalar nonets. The matrices $M$ and $M'$ transform in the same way under $SU(3)_L \times SU(3)_R$ but have different $U(1)_A$ transformation properties.
The Lagrangian has the content:
\begin{eqnarray}
{\cal L}=-\frac{1}{2}{\rm Tr}[D_{\mu}MD^{\mu}M^{\dagger}]-\frac{1}{2}{\rm Tr}[D_{\mu}M^{\prime}D^{\mu}M^{\prime\dagger}]
-V_0(M,M')-V_{SB},
\label{resu56474}
\end{eqnarray}
where,
\begin{eqnarray}
&&D_{\mu}M=\partial_{\mu}M-ieQMA_{\mu}+ieMQA_{\mu}
\nonumber\\
&&D^{\mu}M^{\dagger}=\partial^{\mu}M^{\dagger}+ieM^{\dagger}QA^{\mu}-ieQM^{\dagger}A^{\mu},
\label{cov53628}
\end{eqnarray}
and $Q={\rm diag}(\frac{2}{3}, -\frac{1}{3}, -\frac{1}{3})$.
Here in the leading order of the model which corresponds to retaining only terms with no more than eight quark and antiquark lines,
\begin{eqnarray}
V_0&=&-c_2{\rm Tr}[MM^{\dagger}]+c_4{\rm Tr}[MM^{\dagger}MM^{\dagger}]+d_2{\rm Tr}[M^{\prime}M^{\prime \dagger}]+e_3(\epsilon_{abc}\epsilon^{def}M^a_dM^b_eM^{\prime c}_f+h.c.)+
\nonumber\\
&&c_3\left[\gamma_1\ln\left(\frac{\det M}{\det M^{\dagger}}\right)+(1-\gamma_1)\ln\left(\frac{{\rm Tr}(MM^{\prime\dagger})}{{\rm Tr}(M^{\prime}M^{\dagger})}\right)\right]^2.
\label{potential7356}
\end{eqnarray}
The potential is invariant under $U(3)_L \times U(3)_R$ with the exception of the last term which breaks $U(1)_A$.
The symmetry breaking term has the form:
\begin{eqnarray}
V_{SB}=-2 {\rm Tr}[AS]
\label{sym3528637}
\end{eqnarray}
where $A={\rm diag}(A_1,A_2,A_3)$ is a  matrix proportional to the three light quark masses.
The model allows for two-quark condensates,
$\alpha_a=\langle S_a^a \rangle$ as well as
four-quark condensates
$\beta_a=\langle {S'}_a^a \rangle$.
Here we assume \cite{Schechter1} isotopic spin
symmetry so $A_1$ =$A_2$ and:
\begin{equation}
\alpha_1 = \alpha_2  \ne \alpha_3, \hskip 2cm
\beta_1 = \beta_2  \ne \beta_3
\label{ispinvac}
\end{equation}
We also need the ``minimum" conditions,
\begin{equation}
\left< \frac{\partial V_0}{\partial S}\right> + \left< \frac{\partial
	V_{SB}}{\partial
	S}\right>=0,
\quad \quad \left< \frac{\partial V_0}{\partial S'}\right>
=0.
\label{mincond}
\end{equation}

There are twelve parameters describing the Lagrangian and the
vacuum. These include the six coupling constants
given in Eq.(\ref{potential7356}), the two quark mass parameters,
($A_1=A_2,A_3$) and the four vacuum parameters ($\alpha_1
=\alpha_2,\alpha_3,\beta_1=\beta_2,\beta_3$). The four minimum
equations reduce the number of needed input parameters to
eight.    The details of numerical work for solving this system is given in \cite{Jora5}, and for the readers convenience a summary is given in Appendix A.

The fields of interest are the neutral $I=0$ scalar mesons:

\begin{eqnarray}
&&f_a=\frac{S_1^1+S_2^2}{\sqrt{2}}
\nonumber\\
&&f_b=S^3_3
\nonumber\\
&&f_c=\frac{S^{1\prime}_1+S^{2\prime}_2}{\sqrt{2}}
\nonumber\\
&&f_d=S^{3\prime}_3.
\label{fscalrs546}
\end{eqnarray}
The scalars mix with each other within their group and form the physical states:
\begin{eqnarray}
\left(
\begin{array}{c}
f_1\\
f_2\\
f_3\\
f_4
\end{array}
\right)
=L_0^{-1}
\left(
\begin{array}{c}
f_a\\
f_b\\
f_c\\
f_d
\end{array}
\right)
\label{phys243552}
\end{eqnarray}
Here  $L_0$ is the rotation matrix and depends on the model inputs.
Based on the fit in Ref. \cite{Jora5} the first two physical states are:
\begin{eqnarray}
&&f_1=f_0(500)
\nonumber\\
&&f_2=f_0(980)
\label{eta194760}
\end{eqnarray}
The experimental candidates for the remaining two states predicted by the model ($f_3$ and $f_4$) are $f_0(1370)$, $f_0(1500)$ and $f_0(1710)$.  However, the exact identification requires inclusion of a scalar glueball   which, for simplicity,  was not included in the present order of the model.  In this work our main focus is on $f_0(500)$ and $f_0(980)$.

\section{The trace anomaly term}
The electromagnetic trace anomaly has the expression:
\begin{eqnarray}
\theta^{\mu}_{\mu}=\partial^{\mu}D_{\mu}=-\frac{\beta(e)}{2e}F^{\mu\nu}F_{\mu\nu},
\label{tran967575}
\end{eqnarray}
where $\theta^{\mu}_{\mu}$ is the trace of the energy momentum tensor, $D_{\mu}$ is the dilatation current,  $e$ is the electric charge and $\beta(e)$ is the corresponding beta function. Eq. (\ref{tran967575}) only displays the contribution to trace anomaly due to electromagnetic group which is relevant for the present work.  Note that the full trace anomaly also contains contributions  from the gluon fields and has the expression:
\begin{eqnarray}
\theta^{\mu}_{\mu}=\partial^{\mu}D_{\mu}=-\frac{\beta(e)}{2e}F^{\mu\nu}F_{\mu\nu}-\frac{\beta(g_3)}{2g_3}G^{a\mu\nu}G^a_{\mu\nu},
\label{tran96745575}
\end{eqnarray}
where $g_3$ is the strong coupling constant and $G^{a\mu\nu}$ is the gluon tensor.

 We apply the method introduced in \cite{Schechter1}-\cite{Schechter3} where for an arbitrary Lagrangian with fields $\eta_A$ of mass dimension 1 and $\xi_A$ with mass dimension 4,
 \begin{eqnarray}
 {\cal L}=-\frac{1}{2}\sum_A\partial^{\mu}\eta_A\partial_{\mu}\eta_A-V(\eta_A,\xi_A),
 \label{schlagr3455}
 \end{eqnarray}
 the improved energy momentum tensor is defined as:
 \begin{eqnarray}
 \theta_{\mu\nu}=\delta_{\mu\nu}{\cal L}+\sum_A\partial_{\mu}\eta_A\partial_{\nu}\eta_A-\frac{1}{6}\sum_A(\partial_{\mu}\partial_{\nu}-\delta_{\mu\nu}\Box)\eta_A^2
 \label{thetadef38567}
 \end{eqnarray}
 Here the fields $\eta_A$ and $\xi_A$ transform under the scale transformation:
 \begin{eqnarray}
 &&\delta\eta_A=\eta_A+x_{\mu}\partial^{\mu}\eta_A
 \nonumber\\
 &&\delta\xi_A=4\xi_A+x_{\mu}\partial^{\mu}\xi_A
 \label{tarnsform65774}
 \end{eqnarray}
 The trace the energy momentum tensor can be written then as,
\begin{eqnarray}
\theta^{\mu}_{\mu}=\partial_{\mu}(x^{\mu}{\cal L})-\delta {\cal L},
\label{scatrsnf}
\end{eqnarray}
which can be computed to be,
\begin{eqnarray}
\theta^{\mu}_{\mu}=\sum_A(4\xi_A\frac{\partial V}{\partial \xi_A}+\eta_A\frac{\partial V}{\partial \eta_A})-4V.
\label{finalsc452}
\end{eqnarray}
We shall use the expression in Eq. (\ref{finalsc452}) to derive a suitable effective term that mocks up the electromagnetic anomaly.

Using this approach,  it can be shown that the term
\begin{eqnarray}
{\cal L}_s= bF^{\mu\nu}F_{\mu\nu}
\left\{
\tau_1
\left[
      \ln\left(
      \frac{\det{M}}{\Lambda^3}
        \right)
       +\ln\left(
       \frac{\det{M^{\dagger}}}{\Lambda^3}
       \right)
\right] +
\tau_2
\left[
\ln\left(
\frac{{\rm Tr}MM^{\prime\dagger}}{\Lambda^2}
    \right)
+\ln \left(
\frac{{\rm Tr}M'M^{\dagger}}{\Lambda^2}
      \right)
\right]
\right\}
\label{res645363}
\end{eqnarray}
satisfies the anomaly in Eq.(\ref{tran967575}) provided that the dimensionless coefficients $\tau_1$ and $\tau_2$ satisfy the constraint $6 \tau_1 + 4 \tau_2=1$. Here in calculations the square of the electromagnetic tensor is assimilated to a scalar field of mass dimension 4.
The term in Eq. (\ref{res645363}) is chiral and $U(1)_A$ invariant, constructed by analogy with the axial anomaly and is minimal. It can be however expanded to include other possible contributions with higher orders of  $\Lambda$ (which is expected to be associated with $\Lambda_{QCD}$).
By applying Eq. (\ref{finalsc452}) and requiring that Eq. (\ref{tran967575}) is  satisfied we determine $b=\frac{e^2}{12\pi^2}$ where we used
$\beta(e)=\frac{1}{6\pi^2}e^3$.

In order to determine the coupling of the physical scalars,  we expand the terms in the curly brackets in Eq. (\ref{res645363}) around the vacuum expectation values of $S$ and $S'$ to show that these terms are equal to:
\begin{equation}
\left\{\cdots\right\} =
2\tau_1\left[\frac{1}{\alpha_1}(S^1_1+S^2_2)+\frac{1}{\alpha_3}S^3_3\right]+
\frac{2\tau_2}{2\alpha_1\beta_1+\alpha_3\beta_3}
\left[\alpha_1(S^{1\prime}_1+S^{2\prime}_2)+\alpha_3S^{3\prime}_3+\beta_1(S^1_1+S^2_2)+\beta_3S^3_3\right].
\label{res645353}
\end{equation}
Then the coupling of the physical states with the two photons can be read off easily as:
\begin{equation}
F_{fi} = 4 b \left[
\tau_1 \left(
{{\sqrt 2}\over \alpha_1} (L_0)_{1i} + {1\over \alpha_3} (L_0)_{2i}
\right)+
{\tau_2\over {2 \alpha_1 \beta_1 + \alpha_3 \beta_3}}
\left(
{\sqrt 2} \beta_1 (L_0)_{1i} + \beta_3 (L_0)_{2i}
+ \alpha_1 {\sqrt 2} (L_0)_{3i} + \alpha_3 (L_0)_{4i}
\right)
\right]
\label{coupl7453663}
\end{equation}
where $i=1 \cdots 4$ corresponds to the four isosinglet states (where in this work only the first two are of our interest).  The amplitude of decaying to two photons are:
\begin{eqnarray}
A_i(f_i\rightarrow \gamma\gamma)=-F_{f_i}(k_{1\mu}\epsilon_{1\nu}-k_{1\nu}\epsilon_{1\mu})(k_{2\mu}\epsilon_{2\nu}-k_{2\nu}\epsilon_{2\mu}),
\label{amplit657464}
\end{eqnarray}
where $k_1$, $k_2$ are the photon momenta and $\epsilon_1$, $\epsilon_2$  are the photon polarizations.   The decay width is given by:
\begin{eqnarray}
\Gamma(f_i\rightarrow\gamma\gamma)=F_i^2\frac{m_{f_i}^3}{16\pi}
\label{dec344678}
\end{eqnarray}
where $m_{f_i}$ is the mass of the meson $f_i$.

\section{Decay rates and discussion}

\begin{figure}[!htb]
	\centering
    \includegraphics[scale=0.45]{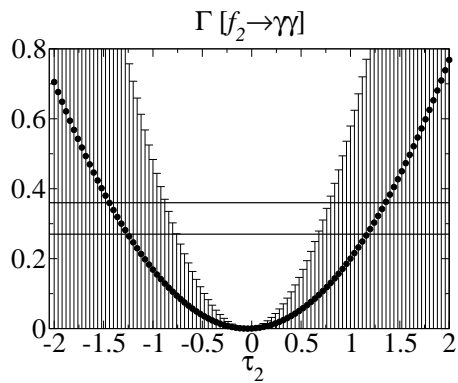}
	\caption{Partial decay width to two photon (KeV) of $f_2$ [or $f_0(980$)] vs $\tau_2$ predicted by the leading order of the generalized linear sigma model.   The error bars represent the uncertainty of the model in its leading order and the circles represent the average predictions at each value of $\tau_2$ (the uncertainties shown stem from the uncertainties of 	$m[\pi(1300)]$ in the range 1.2-1.4 and of the ratio  $A_3/A_1$ varied in the range of 27 to 30).   The two parallel lines show the experimental range for this decay reported in PDG \cite{PDG}.  Overlap with experiment becomes possible  for $\tau_2 \ge 0.7$ and $\tau_2 \le -0.8$.
	}
	\label{F_Gf2_vs_tau2}
\end{figure}

As stated previously, our  focus in this paper is on the two photon decays of $f_0(500)$ and $f_0(980)$.   For $f_0(980)$ the experimental value of $\Gamma\left[f_0(980)\rightarrow\gamma\gamma\right] = 0.31^{+0.05}_{-0.04}$ KeV is listed in PDG \cite{PDG}.  Our model prediction for this decay width is found from  Eqs. (\ref{coupl7453663}) and (\ref{dec344678})  with rotation matrices $L_0$ imported from the prior work \cite{Jora5}. In this estimate the main model uncertainties stem from two of the experimental inputs ($m[\pi(1300)]$ and $A_3/A_1$) used in \cite{Jora5} to fix the model parameters.  In addition, the two new parameters  $\tau_1$ and $\tau_2$
in the trace anomaly in Eq. (\ref{res645363}) are a priori unknown, and therefore, after the constraint $6 \tau_1 + 4 \tau_2 = 1$ is considered, one of them still remains undetermined and needs to be varied (we choose to run $\tau_2$ because it measures the direct effect of chiral nonet mixing on the anomaly term in (\ref{res645363})).
The result is shown in Fig. \ref{F_Gf2_vs_tau2} versus $\tau_2$ with the error bars representing the uncertainties due to variation of $m[\pi(1300)]=1.2-1.4$ and $A_3/A_1=27-30$.   The two horizontal lines give the experimental bounds \cite{PDG} discussed above.   It can be clearly seen that with small $\tau_2$ (which measures the contribution of chiral mixing between nonets $M$ and $M'$) the model predictions do not overlap with the experimental values.   This is very consistent with other observations within this model where it is found that chiral mixing is essential for understanding the global properties of scalar mesons \cite{Jora5}.   It is seen that for values of $\tau_2 \ge 0.7$ and $\tau_2 \le -0.8$ the model predictions overlap with experiment.

\begin{figure}[!htb]
	\centering
	\includegraphics[scale=0.35]{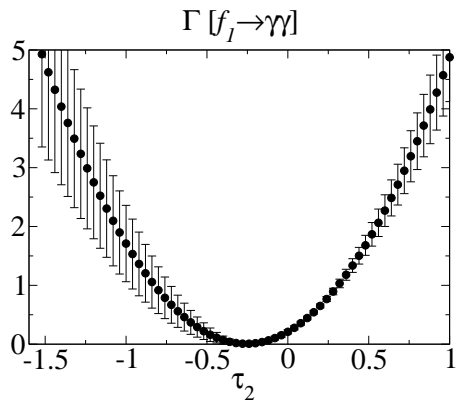}
	\caption{
		Partial decay width to two photon (KeV) of $f_1$ [or $f_0(500$)] vs $\tau_2$ predicted by the leading order of the generalized linear sigma model.   The error bars represent the uncertainty of the model in its leading order and the circles represent the average predictions at each value of $\tau_2$ (the uncertainties shown stem from the uncertainties of 	$m[\pi(1300)]$ in the range 1.2-1.4 and of the ratio  $A_3/A_1$ varied in the range of 27 to 30).
		Comparing with Fig. \ref{F_Gf2_vs_tau2}, the acceptable region is $\tau_2 \ge 0.7$ and $\tau_2 \le -0.8$ which sets a lower bound of approximately 0.5 KeV on the decay with of $f_0(500)$ to two photons.
				 }
	\label{F_Gf1_vs_tau2}
\end{figure}

Similarly, the prediction for the two-photon decay width of $f_0(500)$ is given in Fig. \ref{F_Gf1_vs_tau2}.   Considering the acceptable ranges of $\tau_2 \ge 0.7$ and $\tau_2 \le -0.8$, the prediction of this decay width shows the lower bound of approximately 0.5 KeV (occurring around $\tau_2\approx -0.8 $).   This can be compared with other estimates in the literature such as   $1.2\pm 0.04$ KeV \cite{Bernabeu} or  $10\pm6$ KeV \cite{Courau}. In \cite{Pennington1} the authors made a thorough amplitude analysis of the  experimental data for $\gamma\gamma\rightarrow\pi^+\pi^-$ to find the position of the sigma pole at $0.441-i0.272$ which corresponds to two scenarios for a decay width of $f_0(500)$ to two photons of $3.1\pm 0.5$ KeV or $2.4\pm 0.4$ KeV.  More recently performing a similar analysis Dai and Pennington \cite{Pennington2} found this decay width to be $2.05\pm 0.21$ KeV.    They also found $\Gamma(f_0(980))\rightarrow \gamma\gamma=0.32 \pm0.05$ KeV.
The decay widths of the low lying scalar mesons were analyzed in the literature from the perspective that they proceed mainly through pions and kaons loops \cite{Achasov,Giacosa,Oller,Volkov}.
It is generally hypothesized that because of this assumption it is hard for these decays to be relevant for the quark substructure of the scalar mesons \cite{PDG}.

In summary, the  lower bound of 0.5 KeV for the decay width of $f_0(500)$ to two photons obtained in this analysis (within the leading order of the generalized linear sigma model) is qualitatively consistent with other estimates \cite{Bernabeu,Courau,Pennington1,Pennington2}.   A more precise prediction is expected when higher order effects are taken into account.  Within the current approach it was also shown that chiral nonet mixing is an essential ingredient in understanding this decay width in which direct inclusion of mixing in modeling the trace anomaly is needed.   This last point further supports the importance of underlying mixing among two- and four-quark components in exploring the spectroscopy  of light scalar mesons.

\section*{Acknowledgments} \vskip -.5cm

A.H.F. gratefully  acknowledges the support of College of Arts and Sciences of SUNY Poly in the Fall 2017 semester.

\appendix

\section{Brief review of the Numerical analysis for model parameters and rotation matrices}

In this appendix we give a summary of numerical determination of the eight independent Lagrangian parameters of Eqs. (\ref{resu56474}) and (\ref{potential7356}).   Five of these eight are determined from the following
masses together with the pion decay constant:
\begin{eqnarray}
m[a_0(980)] &=& 980 \pm 20\, {\rm MeV}
\nonumber
\\ m[a_0(1450)] &=& 1474 \pm 19\, {\rm MeV}
\nonumber \\
m[\pi(1300)] &=& 1300 \pm 100\, {\rm MeV}
\nonumber \\
m_\pi &=& 137 \, {\rm MeV}
\nonumber \\
F_\pi &=& 131 \, {\rm MeV}
\label{inputs1}
\end{eqnarray}
Since $m[\pi(1300)]$ has a large uncertainty,
the Lagrangian parameters would depend on
on the choice of this experimental input.
The sixth input is taken as the light
``quark mass ratio" $A_3/A_1$, which are varied over its appropriate range (in this work we use 27-30).

The remaining two parameters ($c_3$ and $\gamma_1$) only affect the isosinglet pseudoscalars (whose properties also
depend on the ten parameters discussed above).    However, there are several choices for determination of these two parameters depending on how the   four isosinglet pseudoscalars predicted in this model are matched to many experimental candidates below 2 GeV.   The two lightest predicted by the model ($\eta_1$ and $\eta_2$)  are identified with $\eta(547)$ and $\eta'(958)$ with masses:
\begin{eqnarray}
m^{\rm exp.}[\eta (547)] &=& 547.853 \pm
0.024\, {\rm
	MeV},\nonumber \\
m^{\rm exp.}[\eta' (958)] &=& 957.78 \pm 0.06
\, {\rm
	MeV}.
\end{eqnarray}
For the two heavier ones ($\eta_3$ and $\eta_4$),   there are six ways that they can be identified with the four experimental candidates above 1 GeV:  $\eta(1295)$,  $\eta(1405)$,  $\eta(1475)$, and $\eta(1760)$ with masses,
\begin{eqnarray}
m^{\rm exp.}[\eta (1295)] &=& 1294 \pm 4\, {\rm
	MeV},\nonumber \\
m^{\rm exp.}[\eta (1405)] &=& 1409.8 \pm 2.4 \,
{\rm
	MeV},
\nonumber \\
m^{\rm exp.}[\eta (1475)] &=& 1476 \pm 4\, {\rm
	MeV},\nonumber \\
m^{\rm exp.}[\eta (1760)] &=& 1756 \pm 9 \,
{\rm
	MeV}.
\end{eqnarray}
This leads to six scenarios considered in detail in \cite{Jora5}.
The two experimental inputs for determination of the two parameters $c_3$ and $\gamma_1$ are taken to be  Tr$M_\eta^2$ and det$M_\eta^2$, i.e.
\begin{eqnarray}
{\rm Tr}\, \left(  M^2_\eta  \right) &=&
{\rm Tr}\, \left(  {M^2_\eta}  \right)_{\rm exp},
\nonumber \\
{\rm det}\, \left( M^2_\eta \right) &=&
{\rm det}\, \left( {M^2_\eta} \right)_{\rm exp}.
\label{trace_det_eq}
\end{eqnarray}
Moreover,  for each of the six scenarios,  $\gamma_1$ is found from a quadratic equation, and as a result, there are altogether twelve possibilities for determination of $\gamma_1$ and $c_3$.    Since only Tr and det of experimental masses are imposed for each of these twelve possibilities, the resulting  $\gamma_1$ and $c_3$ do not necessarily recover the exact individual experimental masses,  therefore the best overall agreement between the predicted masses (for each of the twelve possibilities) were examined in \cite{Jora5}.   Quantitatively,  the
goodness of each solution was measured by the smallness of
the following quantity:
\begin{equation}
\chi_{sl} =
\sum_{k=1}^4
{
	{\left| m^{\rm theo.}_{sl}(\eta_k)  -
		m^{\rm exp.}_{s}(\eta_k)\right|}
	\over
	m^{\rm exp.}_{s}(\eta_k)
},
\label{E_chi_sl}
\end{equation}
in which $s$ corresponds to the scenario
(i.e. $s= 1 \cdots 6$) and
$l$ corresponds to the solution number
(i.e. $l=$ I, II).   The quantity $\chi_{sl}
\times 100$ gives the overall percent
discrepancy between our theoretical prediction
and experiment.   For the six scenarios and
the two solutions for each scenario,
$\chi_{sl}$ was analyzed  in ref. \cite{Jora5}.
For the third scenario (corresponding to identification of $\eta_3$ and $\eta_4$ with experimental candidates $\eta(1295)$ and $\eta(1760)$) and  solution I the best agreement with the mass spectrum of the eta system was obtained (i.e. $\chi_{3\rm{I}}$ was the smallest).      Furthermore,   all six scenarios were examined in the analysis of $\eta'\rightarrow\eta\pi\pi$ decay in \cite{14_FSZZ} and it was found that the best overall result (both for the partial decay width of $\eta'\rightarrow \eta\pi\pi$ as well as the energy dependence of its squared decay amplitude) is obtained for scenario ``3I'' consistent with the analysis of ref. \cite{Jora5}. In this work,  we use the result of ``3I'' scenario.

The numerical values for the rotation matrix $L_0$ defined in (\ref{coupl7453663}) can be consequently determined.   Since two of the model inputs $A_3/A_1$ and $m[\pi(1300)]$ have large uncertainties, the numerical values of these rotation matrices naturally have some dependencies on these two inputs.   Table \ref{L_num} gives numerical values of $L_0$ for three values of $m[\pi(1300)]$ and three values of $A_3/A_1$.

\begin{table}[!htbp]
	\centering
	\caption{
		Rotation matrix $L_0$ for several values of  $A_3/A_1$ and $m[\pi(1300)]$.
	}
	\renewcommand{\tabcolsep}{0.4pc} 
	\renewcommand{\arraystretch}{1.5} 
	\begin{tabular}{c||c|c|c}
		\noalign{\hrule height 1pt}
		\noalign{\hrule height 1pt}
		$\begin{array}{c}
		m[\pi(1300)] ({\rm GeV}) \rightarrow\\
		A_3/A_1 \downarrow
		\end{array}$		& 1.2 & 1.3 & 1.4   \\
		\noalign{\hrule height 1pt}
		27.0 &
		$\begin{array}{cccc}
		0.586 & -0.110 & 0.800 & 0.065 \\
		0.204 & 0.192 & -0.045 &-0.959\\
		0.608 & 0.641 & -0.380 & 0.275\\
		0.496 & -0.735& -0.462 & -0.020
		\end{array}$
		&	
		$\begin{array}{cccc}
	   -0.678 & 0.159 & -0.717 & -0.001\\
	   -0.246 & -0.284 & 0.169 & 0.911\\
	   -0.543 & -0.629 & 0.374 & -0.412\\
	   -0.430 & 0.706 & 0.563 & 0.000
		\end{array}$	 		
		&
		$\begin{array}{cccc}
		0.778 & -0.233 & -0.557 & 0.172\\
		0.240 &  0.467 & 0.376  & 0.764\\
		0.451 & 0.625 & 0.179  & -0.612\\
		0.366 & -0.580 & 0.719 & -0.114
		\end{array}$
		\\
		\hline
		28.5 &
		$\begin{array}{cccc}
		0.585 & -0.105 & 0.801 & 0.065\\
		0.198 & 0.185  &-0.042 &-0.962\\
		0.609 & 0.643  & -0.382 & 0.266\\
		0.498 & -0.736 & -0.459 & -0.019
		\end{array}$
        &
		$\begin{array}{cccc}
		-0.679 & 0.152 & 0.718 & 0.008\\
		-0.239 & -0.274 &-0.158 & -0.918\\
		-0.544 & -0.632 & -0.384 & 0.396\\
		-0.432 & 0.708 & -0.558 & -0.003
		\end{array}$
		&
		$\begin{array}{cccc}
	    0.780 & -0.223 & -0.564 & 0.154\\
	    0.232 & 0.449  & 0.359  & 0.785\\
	    0.450 & 0.634 & 0.210 & -0.592\\
	    0.366 & -0.589 & 0.714 & -0.098
		\end{array}$
		\\
		\hline
		30.0 &
		$\begin{array}{cccc}
		0.585 & -0.101 & 0.802 & 0.065\\
		0.192 & 0.179 & -0.040 & -0.964\\
		0.610 & 0.644 & -0.384 & 0.257\\
		0.499 & -0.737 & -0.456 & -0.018
		\end{array}$
		&
		$\begin{array}{cccc}
		0.680 & -0.146 & 0.718 & 0.014\\
		0.232 & 0.264 & -0.148 & -0.924\\
		0.544 & 0.635 & -0.393 & 0.381\\
		0.433 & -0.711 & -0.554 & -0.006
		\end{array}$
		&
		$\begin{array}{cccc}
		0.782 & -0.214 & -0.569 & 0.138\\
		0.225 & 0.431 & 0.343 & 0.804\\
		0.450 & 0.642 & 0.239 & -0.572\\
		0.367 & -0.597 & 0.708 & -0.085
		\end{array}$\\
		\hline
	\end{tabular}\\[2pt]
	\label{L_num}
\end{table}

\end{document}